\documentclass{PoS}
\usepackage{epsfig}

\newcommand{\Tr}{\mbox{Tr}}
\def\bra#1#2{\ifx#2\ket\langle#1\else\langle#1\vert\fi#2}
\def\ket#1{\vert#1\rangle}

\leftline{\parbox{3cm}{\large\rm DESY 07-164 \\HU-EP-07/33\\LMU-ASC 55/07}
\vspace{1cm}}

\title{The chiral transition on a $24^3\times 10$ lattice with $N_f=2$ clover sea quarks studied by overlap valence quarks}

\ShortTitle{The chiral transition  on a $24^3\times 10$ lattice  studied by overlap valence quarks}

\author{\speaker{Volker Weinberg}$^{\hspace{1mm}a,b}$, Ernst-Michael Ilgenfritz$^{c}$, Karl Koller$^{d}$, Yoshiaki Koma$^{e}$, Yoshifumi Nakamura$^{a}$, Gerrit Schierholz$^{a,f}$, Thomas Streuer$^{g}$\\
~\\

$^{a}$ John von Neumann-Institut f\"ur Computing NIC, 15738 Zeuthen, Germany\\
$^{b}$ Institut f\"ur theoretische Physik, Freie Universit\"at Berlin, 14196 Berlin, Germany\\
     $^{c}$ Institut f\"ur Physik, Humboldt Universit\"at zu Berlin, 12489 Berlin, Germany\\
$^{d}$ Sektion Physik, Universit\"at M\"unchen, 80333 M\"unchen, Germany\\
$^{e}$ Numazu National College of Technology, Numazu 410-8501, Japan\\
$^{f}$ Deutsches Elektronen-Synchrotron DESY, 22603 Hamburg, Germany\\
$^{g}$ Dep. of Physics and Astronomy, University of Kentucky, Lexington, 
KY 40506-0055, USA 
}
\author{\centerline{\rm E-Mail: \it volker.weinberg@desy.de}}

\author{\centerline{\large \sc For the DIK-collaboration}}

\abstract{
Overlap fermions are particularly well suited to study the finite temperature  dynamics of the 
chiral symmetry restoration transition of QCD, which  might be just an analytic crossover.
Using gauge field configurations on a $24^3\times10$ lattice with $N_f=2$ flavours 
of dynamical Wilson-clover quarks generated by the DIK collaboration, we compute 
the lowest 50 eigenmodes of the overlap Dirac operator and try to locate the transition  by fermionic means. We analyse the spectral
density, local chirality and localisation properties of the low-lying modes 
and illustrate the changing topological and 
(anti-) selfdual structure of the underlying gauge fields across the transition.

}

\FullConference{The XXV International Symposium on Lattice Field Theory\\
                 July 30 - August 4, 2007\\
                 Regensburg, Germany}

\begin{document}

\section{Introduction and simulation parameters}

Despite  enormous theoretical efforts the very nature and order of the QCD 
finite temperature transition is still under debate and subject of 
current research (see e.g. \cite{fodorandkarschtalk} at this conference).
To investigate confinement related  aspects of the transition, during the last six years 
the DIK collaboration has generated dynamical configurations with $N_f=2$ 
flavours of $O(a)$ improved Wilson sea quarks on a $16^3\times 8$ lattice
at $\beta=5.2$ and 5.25~\cite{Bornyakov:2004ii}, a  $24^3\times 10$ lattice
at $\beta=5.20$~\cite{Bornyakov:2005dt} and recently also a $24^3\times 12$ 
lattice at $\beta=5.29$~\cite{bornyakovtalk}. Based on the Polyakov loop 
susceptibility the critical temperature extrapolated to the continuum limit 
at physical $m_\pi$ is determined as $r_0T_c = 0.438(6)(^{+13}_{-7})$~\cite{bornyakovtalk}.

Around three years ago we began to 
include the  chiral symmetry breaking/restoration aspects of the QCD finite 
temperature transition to the topics of interest.
Since overlap fermions implement exact chiral symmetry and the index theorem 
on the lattice, they are specially suited to study various chiral symmetry 
and topology related properties of the  transition.   
First results using valence overlap fermions as a probe for dynamical 
$16^3\times 8$ configurations at $\beta=5.2$ were reported at 
LATTICE 2005~\cite{Weinberg:2005dh}. 
Meanwhile, we have developed a couple of tools based on the overlap Dirac 
operator and its eigenmodes and 
learned to use them for the investigation of the vacuum
structure of quenched QCD 
at $T=0$~\footnote{See~\cite{Ilgenfritz:2007xu} for a more detailed 
description of our tools and methods used throughout this paper.}. 

In this paper we present results obtained by applying these 
methods in a hybrid approach to the $24^3\times 10$ dynamical DIK 
configurations at finite $T$. We try to work out those signals  
which exhibit a remarkable 
difference between the low- and high-temperature phase of QCD.
Using the Arnoldi algorithm we have computed the 50 lowest eigenvalues 
$i\lambda_i$ and eigenvectors $\ket{\psi_i(x)}$ (normalised as 
$\bra{\psi_i}\ket{\psi_i}=1$) of the massless improved overlap operator 
$D(m_q=0)$ for seven ensembles with $\kappa$ values in the vicinity of 
the transition region. In~\cite{Bornyakov:2005dt} the critical value of $\kappa$ marking the transition, $\kappa_t$, for this set of lattice configurations at fixed $\beta=5.2$ 
has been determined from the peak of the Polyakov loop susceptibility 
$\chi_L$ shown in Fig.~\ref{fig:suscpt} (a) as $\kappa_t=0.13542(6)$ and 
assigned to $r_0 T_c=0.499(5)$ 
using interpolated QCDSF values for the Sommer parameter $r_0/a$ found at 
$T=0$. The number of configurations used in our overlap analysis together 
with the values for $T/T_c$ and $r_0/a$ are shown in 
Table~\ref{simulationparameters}.  

\begin{table}[h!t]
\begin{center}

\begin{tabular}{|c|c|c|c|c|}\hline
 $\kappa$ & \# confs& $T/T_c$& $r_0/a$ \\\hline\hline
 0.1348  & 131  & 0.91 & 4.561 \\
 0.1352  & 86   & 0.97 & 4.832\\
 0.1353  & 131  & 0.98 & 4.902\\
 0.1354  & 97   & 1.00 & 4.973\\%\hline\hline
 0.1355  & 118  & 1.01 & 5.045\\
 0.1358  & 122  & 1.06 & 5.265\\
 0.1360  & 97   & 1.09 & 5.417\\\hline
\end{tabular}

\end{center}
\label{simulationparameters}
\caption{Simulation parameters for the configurations on the $24^3\times 10$ lattice generated at $\beta=5.2$.}
\end{table}

\section{Fermionic spectral approaches to locate the transition}

\begin{figure}[ht]
\hspace*{-0.5cm}\begin{tabular}{cc}
\epsfig{file=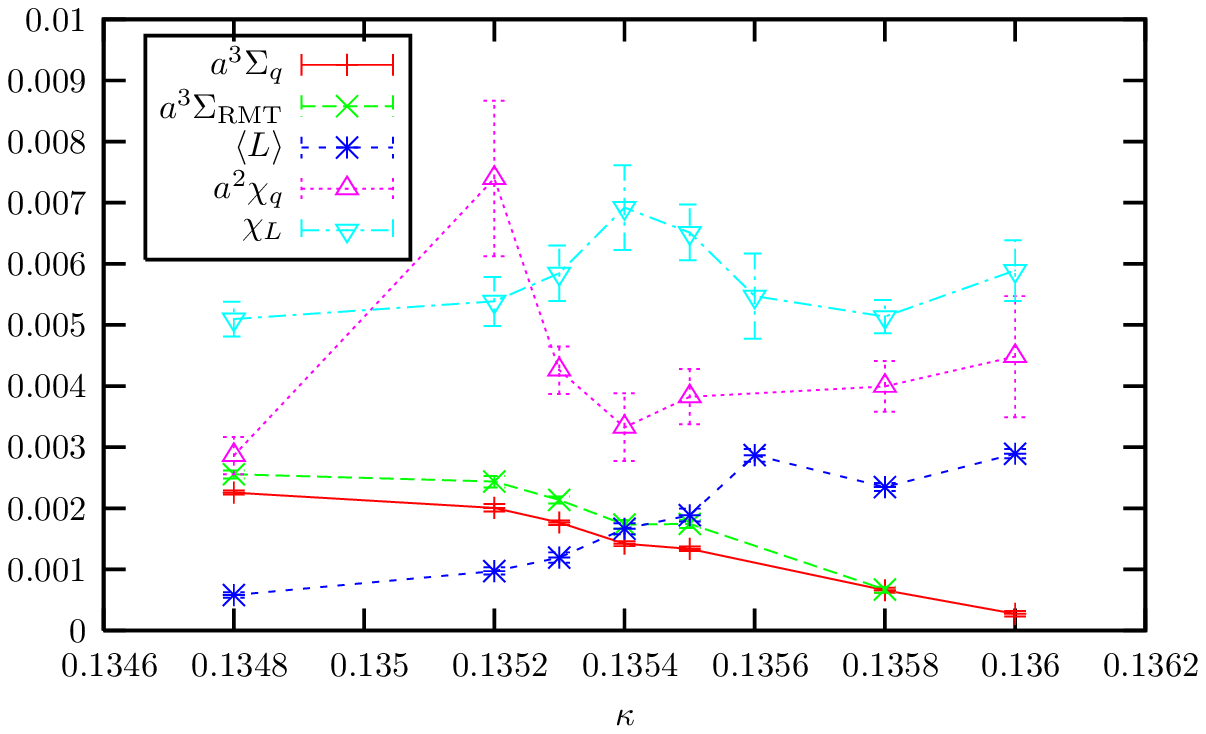,width=7.4cm,clip=}&
\epsfig{file=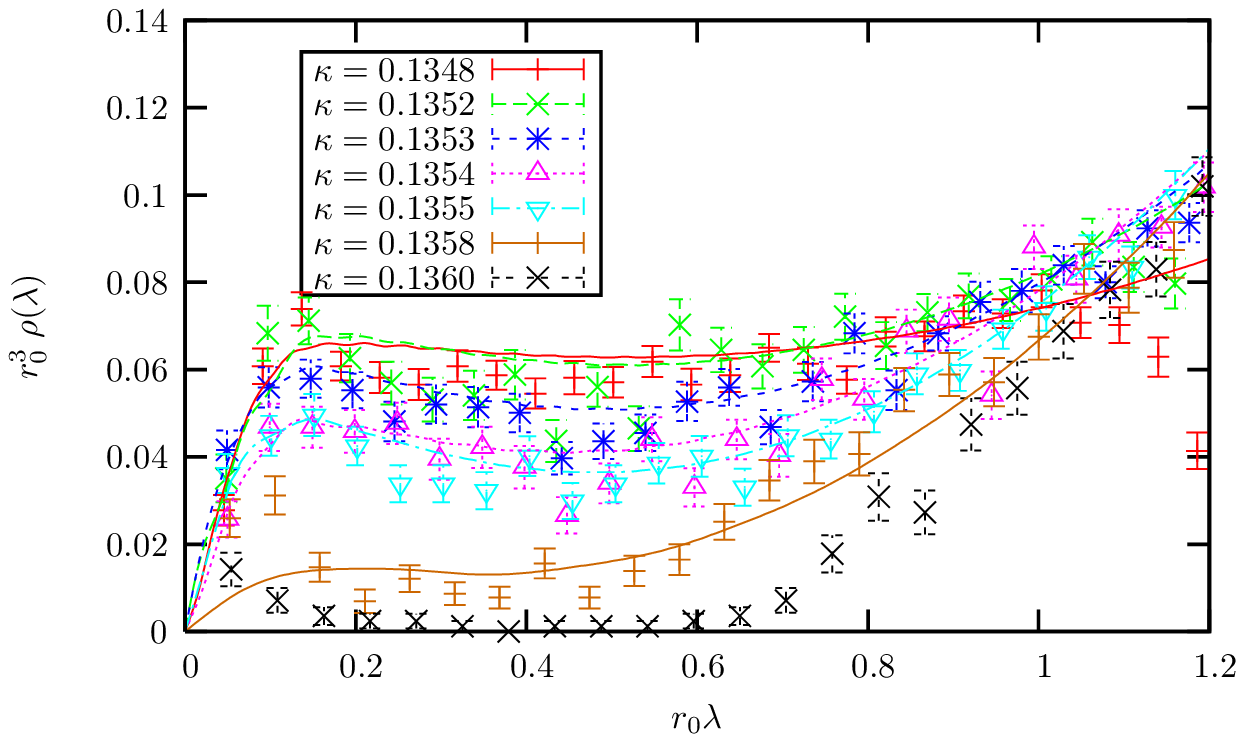,height=4.5cm}\\
(a) & (b)
\end{tabular}
\caption{ (a) The Polyakov loop $\langle L\rangle$ and its susceptibility 
$\chi_L$ shown together with the  chiral condensates $\Sigma_{q,{\rm RMT}}$ and 
the chiral susceptibility $\chi_q$. $\Sigma_q(m_q)$ and $\chi_q$ are 
computed from the spectral representation of the chiral condensate using 
50 eigenmodes, while $\Sigma_{\rm RMT}$ is obtained from the fits of the 
spectral density shown on the right. 
The two chiral condensates $\Sigma_q$ and $\Sigma_{\rm RMT}$ agree quite well  
as functions of $\kappa$ without further rescaling. To present 
the other curves in the same picture, they have been scaled  appropriately.
(b) The spectral densities $\rho(\lambda)$ together with fits using quenched 
random matrix theory predictions.
}
\label{fig:suscpt}
\end{figure}

To determine the critical value of $\kappa$ by fermionic means, we calculate the 
disconnected part of the chiral susceptibility 
$\chi_q=1/V \;(\langle(\Tr\; D^{-1}(m_q))^2\rangle
             -\langle\Tr\; D^{-1}(m_q)\rangle^2)$ %~\cite{Cheng:2006qk}  
using a spectral decomposition of the chiral condensate 
$\Sigma_q(m_q)=1/V\; \langle\Tr\; D^{-1}(m_q)\rangle=1/V\;\langle\sum_i 1/(i\lambda_i+m_q)\rangle$. 
Truncating the decomposition acts as an UV-filter by removing short-distance 
fluctuations
from the local condensate.
We match the overlap valence and the Wilson sea quark
masses by demanding that the corresponding pion masses be equal. The
resulting quark masses range from $am_q=0.045$ at $\kappa=0.1348$ to
$am_q=0.006$ at $\kappa=0.1360$.
In Fig.~\ref{fig:suscpt} (a) one can see that $\chi_q$ shows a peak at 
$\kappa \approx 0.1352$. This value is significantly lower than 
$\kappa_t = 0.13542(6)$ as determined by the Polyakov loop susceptibility 
$\chi_L$. 

According to the Banks-Casher relation $\Sigma=-\pi\rho(0)$ the 
appearance of a gap in the spectrum is a criterion for a chiral symmetry
restoring phase transition. We show in Fig.~\ref{fig:suscpt} (b) the spectral 
density $\rho(\lambda)$ of nonzero modes among the 50 lowest overlap operator 
eigenmodes for the seven analysed ensembles 
together with fits using quenched random matrix theory.\footnote{Since the general formulae of RMT for $N_f=2$ flavours converge to the quenched expressions for large quark masses, quenched RMT can be used as an approximation in the confined phase.}
One can see that a gap in the spectrum does not appear below 
$\kappa\approx 0.1358$. As we have already observed on the $16^3\times 8$ 
lattice~\cite{Weinberg:2005dh}, even for the highest analysed $\kappa$ value 
some eigenvalues, which in fact exclusively belong to the first pair of 
nonzero modes, fall into the would-be gap. 

\begin{figure}[ht]
\begin{center}
\epsfig{file=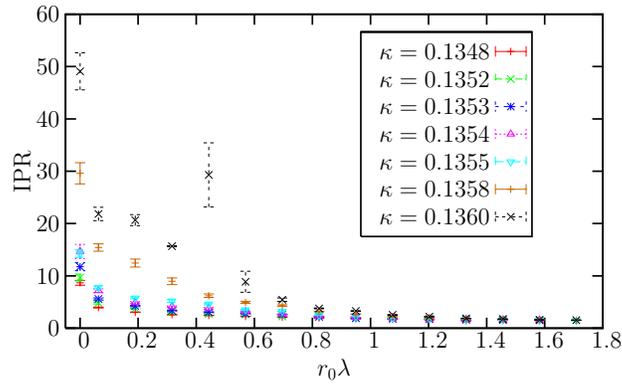,height=5cm}
\caption{The average IPR for zero modes and for nonzero modes in bins of 
$r_0\lambda$ with a width 0.125.}
\label{fig:rho-ipr}
\end{center}
\end{figure}

Due to the hypothetic pinning-down of the lowest modes on singular confining 
defects~\cite{ilgenfritztalk} and their changing structure it is tempting 
to investigate the localisation properties of the low-lying modes.
In Fig.~\ref{fig:rho-ipr} we plot for all considered ensembles the average 
Inverse Participation Ratio (IPR)~ $I=V\sum_x p_i(x)^2$, with the scalar density 
$p_i(x)=\bra{\psi_i(x)}\ket{\psi_i(x)}$, 
for $\lambda_i$ in the respective bin. While the higher modes in 
the bulk of the spectrum are delocalised ({\it i.e.} have small IPR) in both 
phases, the zero modes and low-lying modes are strongly localised ({\it i.e.} 
have large IPR) in the confined phase. The transition is preceded by 
the lowest modes becoming even more localised before a gap finally opens. For 
the two largest $\kappa$-values the isolated modes which fall into the gap  
are extremely localised.

\begin{figure}[b]
\begin{center}
\hspace*{-2cm}\begin{tabular}{cc|cc}
\includegraphics[width=5cm]{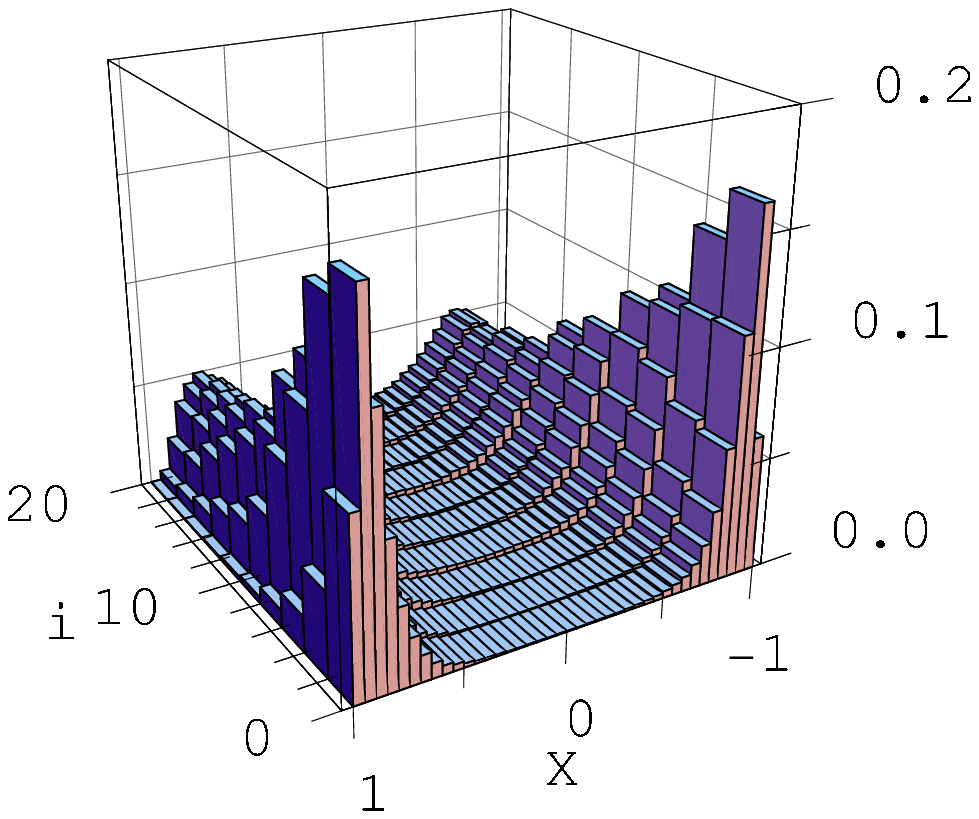}&\hspace*{-1.3cm}\includegraphics[width=5cm]{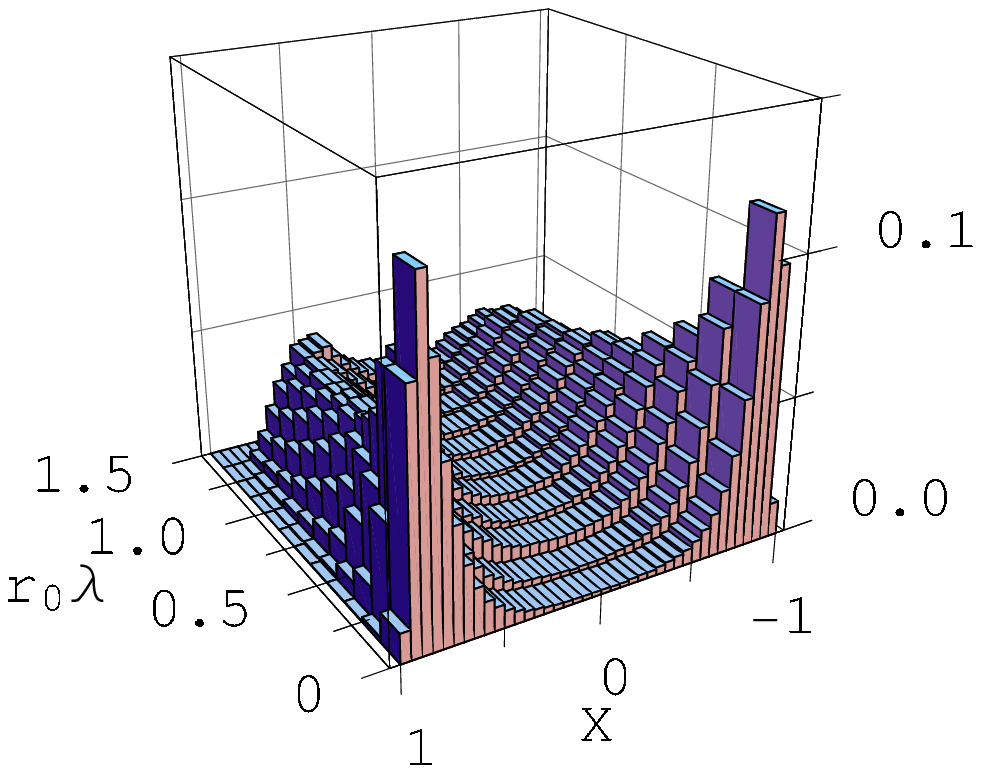}&
\hspace*{-0.8cm}\includegraphics[width=5cm]{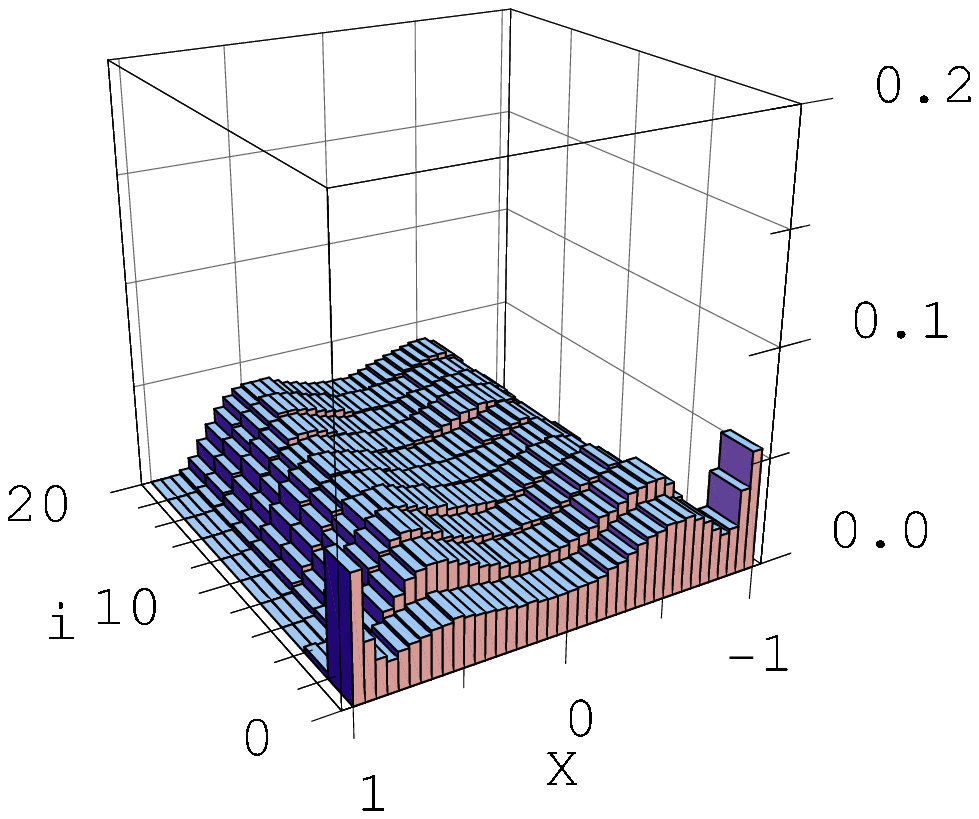} &\hspace*{-1.3cm}\includegraphics[width=5cm]{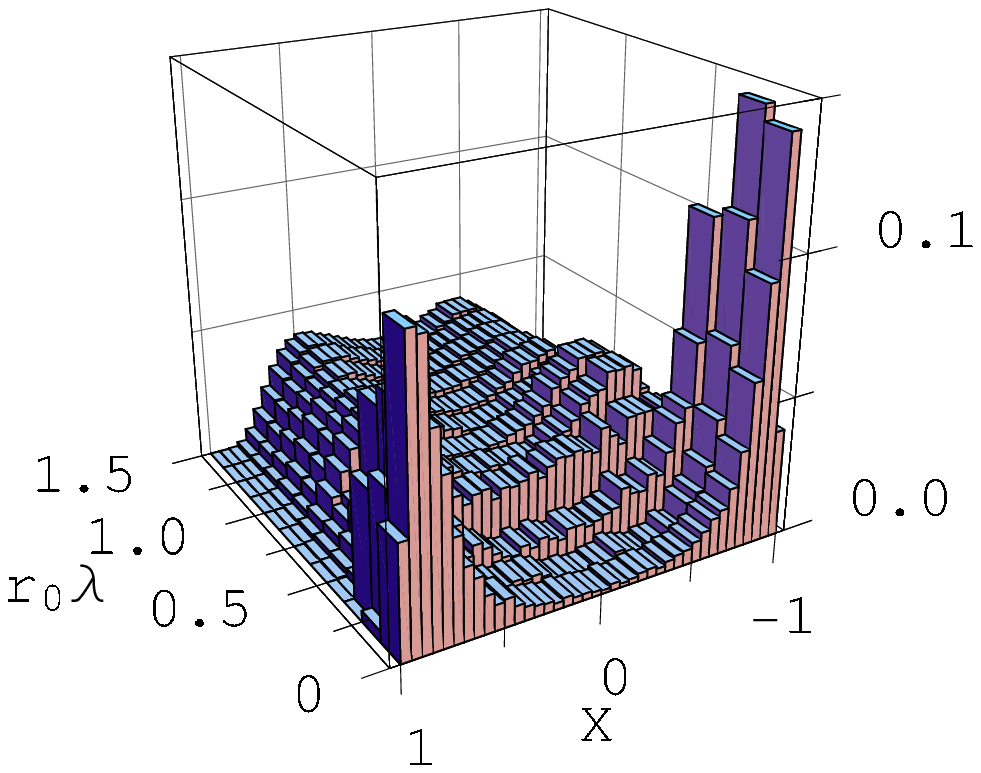} \\[-1cm]
\multicolumn{2}{c}{$\kappa=0.1352$} &\multicolumn{2}{c}{$\kappa=0.1360$}
\end{tabular}
\end{center}
\caption{Normalised histograms of the local chirality $X(x)$ averaged over all 
configurations for $\kappa=0.1352$ (left) and  $\kappa=0.1360$ (right). 
In both cases the left subpanel shows the chirality $X(x)$ for the lowest 
$i=1,\dots,20$ nonzero modes, whereas the right subpanel shows it for all nonzero 
modes averaged over bins in $r_0\lambda$ with bin width 0.125.
Only 1 \% of the lattice sites with largest scalar density $p(x)$ are considered.}
\label{fig:localchir}
\end{figure}

While the zero modes of the overlap operator are exactly chiral, {\it i.e.} 
$p_{5~i}(x):=\bra{\psi_i(x)}\gamma_5\ket{\psi_i(x)}$\\
$=\pm p_i(x)$, the nonzero 
modes have globally vanishing chirality, $\sum_x p_{5~i}(x)=0$, but still exhibit 
a rich local chirality structure correlated with the underlying gauge fields. 
To visualise the changes of the local chirality of the nonzero modes in the 
vicinity of the transition, in Fig.~\ref{fig:localchir} we show histograms 
of the local chirality variable characterising a mode at $x$,  
$X(x)=\frac{4}{\pi}\arctan\left(\sqrt{\frac{p_+(x)}{p_-(x)}}  \right)-1$ with  
$p_{\pm~i}(x)=\bra{\psi_i(x)}\frac{1}{2}(1\pm\gamma_5) \ket{\psi_i(x)}$ introduced 
in~\cite{Horvath:2001ir}.  
$X(x)$ clusters near $\pm 1$ in the confined phase for the low modes when one 
selects lattice points near the peaks of the scalar density $p(x)$, signalling 
a high amount of local chirality. As $\lambda$ increases, the signal for local 
chirality weakens. On the other hand, in the high-temperature phase the signal 
completely vanishes for modes outside the spectral gap. Only the modes which 
fall into the gap show a remanent strong local chirality.

\begin{figure}[ht]
\begin{center}
\hspace*{-0.6cm}\begin{tabular}{cc}
\includegraphics[height=7.5cm,angle=-90]{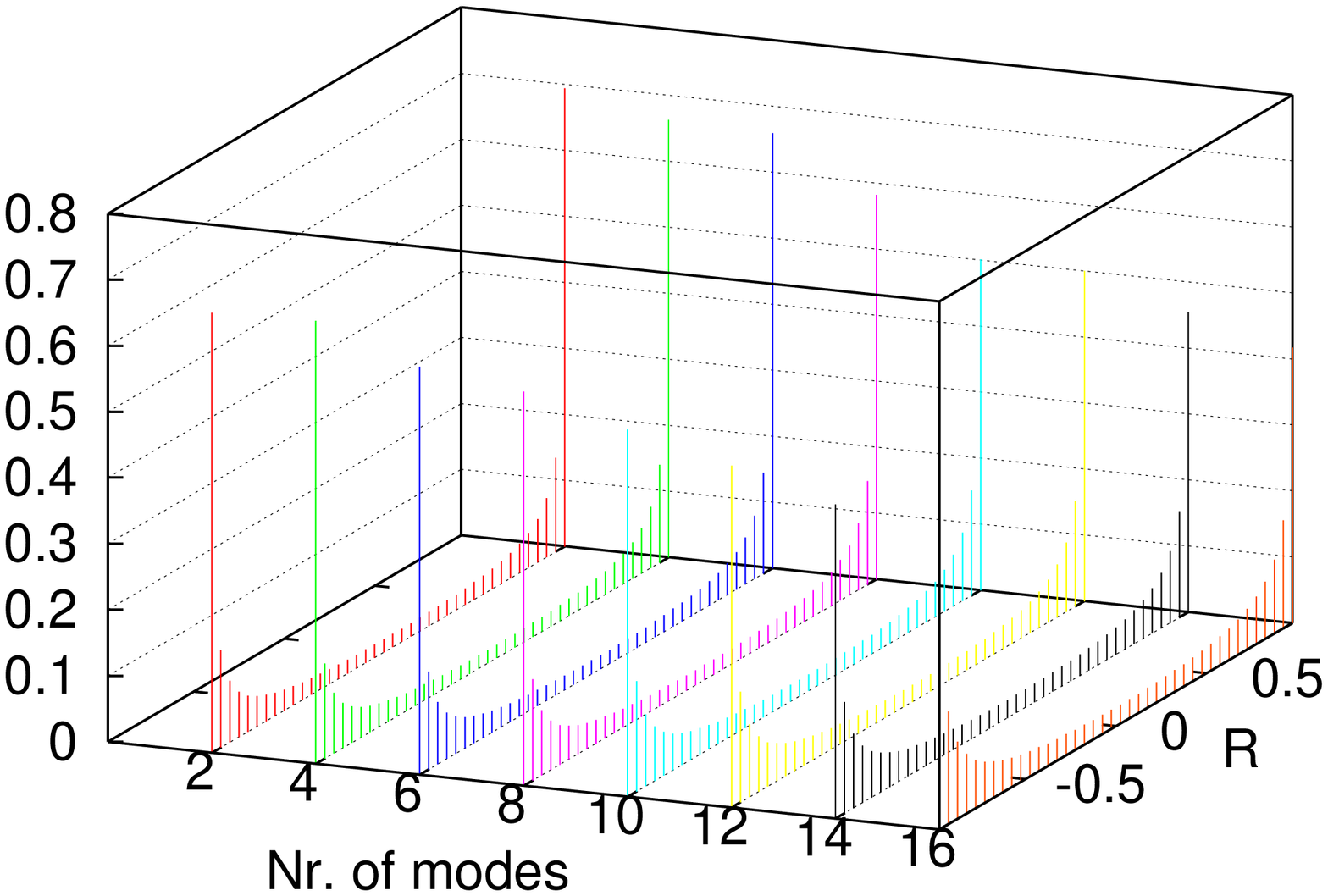} &\includegraphics[height=7.5cm,angle=-90]{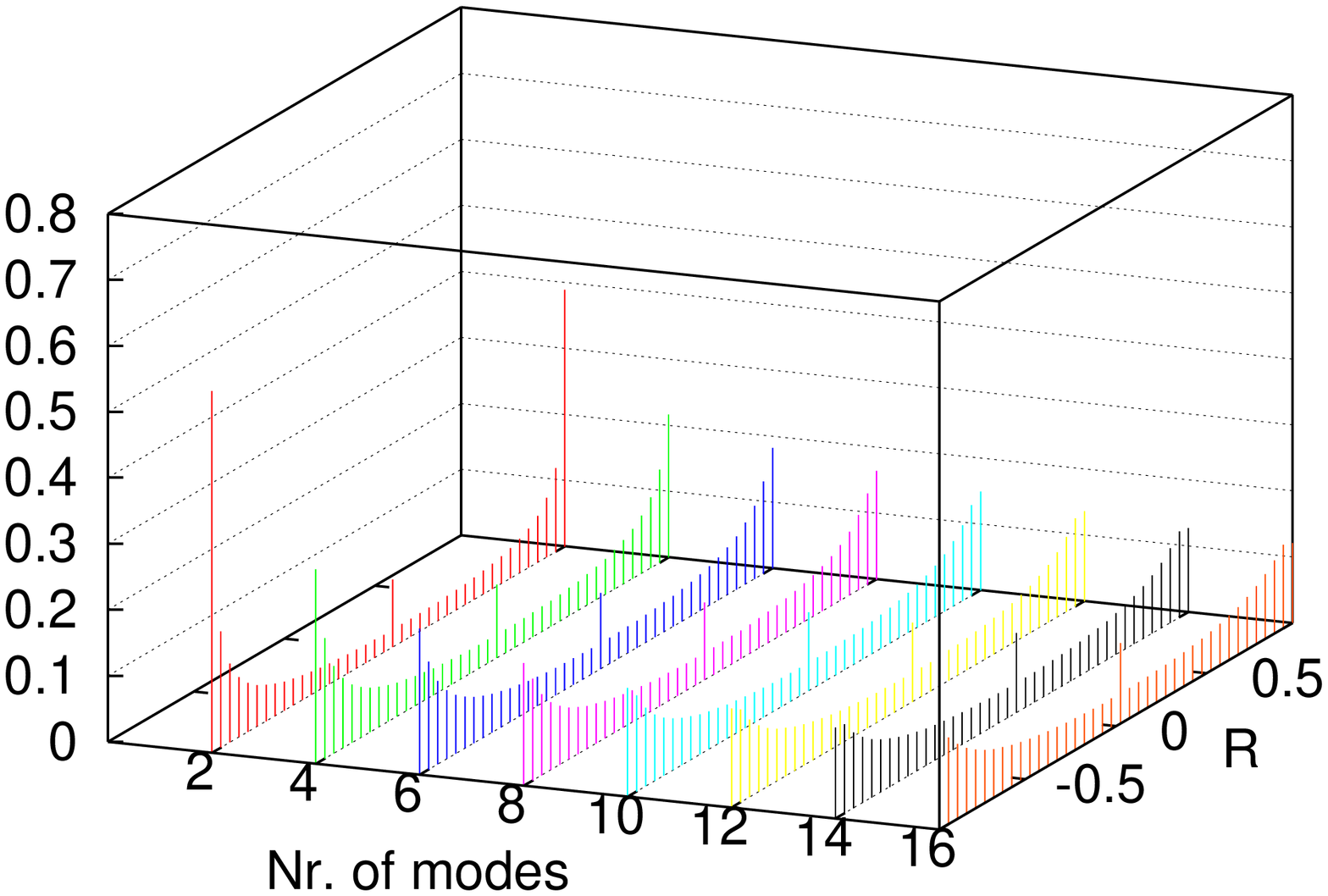}\\[-0.5cm]
$\kappa=0.1352$ & $\kappa=0.1360$
\end{tabular}
\end{center}
\caption{Normalised histograms with respect to the local (anti-)selfduality $X(x)$ 
of the UV-filtered field strength tensor in the $Q=0$ subsample for 
$\kappa=0.1352$ (left) and $\kappa=0.1360$ (right) taken over all lattice sites 
in dependence of the number of nonzero modes included in the ``filter''.}
\label{fig:selfduality} 

\end{figure}
\newpage

Similar changes happen with the distribution of 
$R(x) = \frac{4}{\pi}\arctan r(x) - 1$, with 
$r(x)=(\tilde s(x)-\tilde q(x))/(\tilde s(x)+\tilde q(x)) \;$, 
a measure proposed by Gattringer~\cite{Gattringer} to describe the local degree 
of (anti-) selfduality of the gluonic field strength tensor. 
Using a spectral decomposition of the gluonic field strength tensor, an 
UV-filtered version of the action density 
$\tilde s(x)$ = $\sum_{i,j = 1}^n \frac{\lambda_i^2 \lambda_j^2}{2} f^a_{\mu \nu}(x)_i f^a_{\mu \nu}(x)_j$ 
and of the charge density  
$\tilde q(x)$ = $\sum_{i,j = 1}^n \frac{\lambda_i^2 \lambda_j^2}{2} f^a_{\mu \nu}(x)_i \widetilde{f^a_{\mu \nu}(x)_j}$ 
~(with $f^a_{\mu \nu}(x)_i  =  -\frac{i}{2} \; \bra{\psi_i(x)} \gamma_\mu \gamma_\nu T^a\ket{\psi_i(x)}$) 
can be obtained from the overlap eigenmodes.
$R(x)$ clusters near  (+1) -1 for approximately (anti-)self\-dual fields.
In Fig.~\ref{fig:selfduality} we show that the contribution of the lowest modes 
to the spectral decomposition of the UV-filtered gluonic field strength tensor 
is highly (anti-)selfdual in the low-temperature phase, whereas in the 
high-temperature phase the coherence in the spectral decomposition  
which is necessary to build up an (anti-)selfdual UV-filtered field strength is almost
completely lost.
\begin{center}
\begin{figure}[ht]
\hspace*{-0cm}\begin{tabular}{cc}
\includegraphics[height=3.8cm]{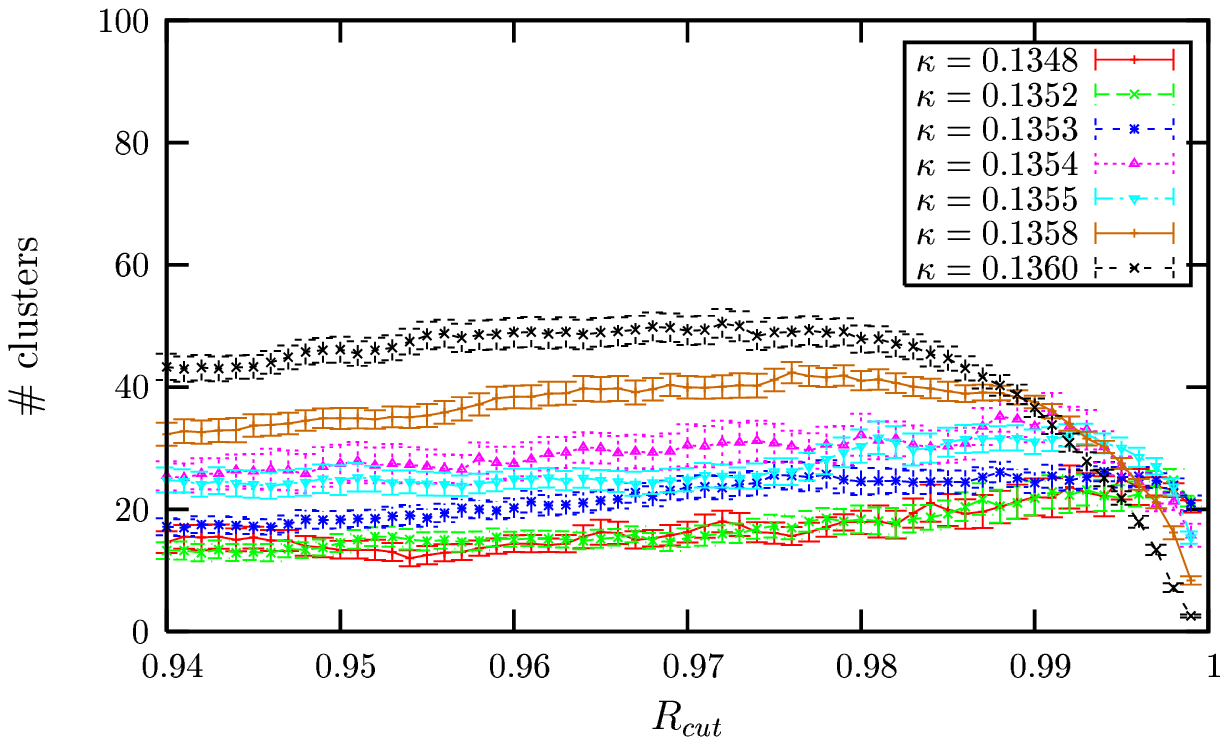}&
\includegraphics[height=3.8cm]{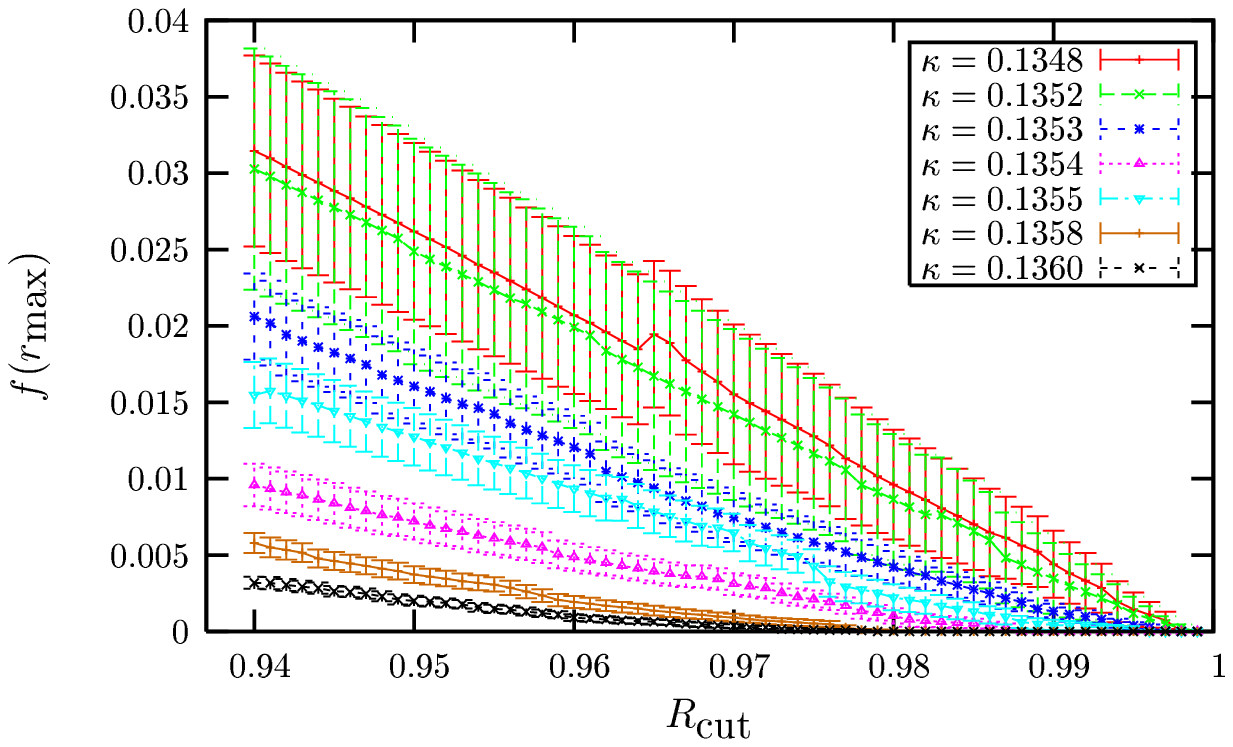}\\
(a) & (b) \\
\hspace*{-0.7cm}\includegraphics[height=3.8cm]{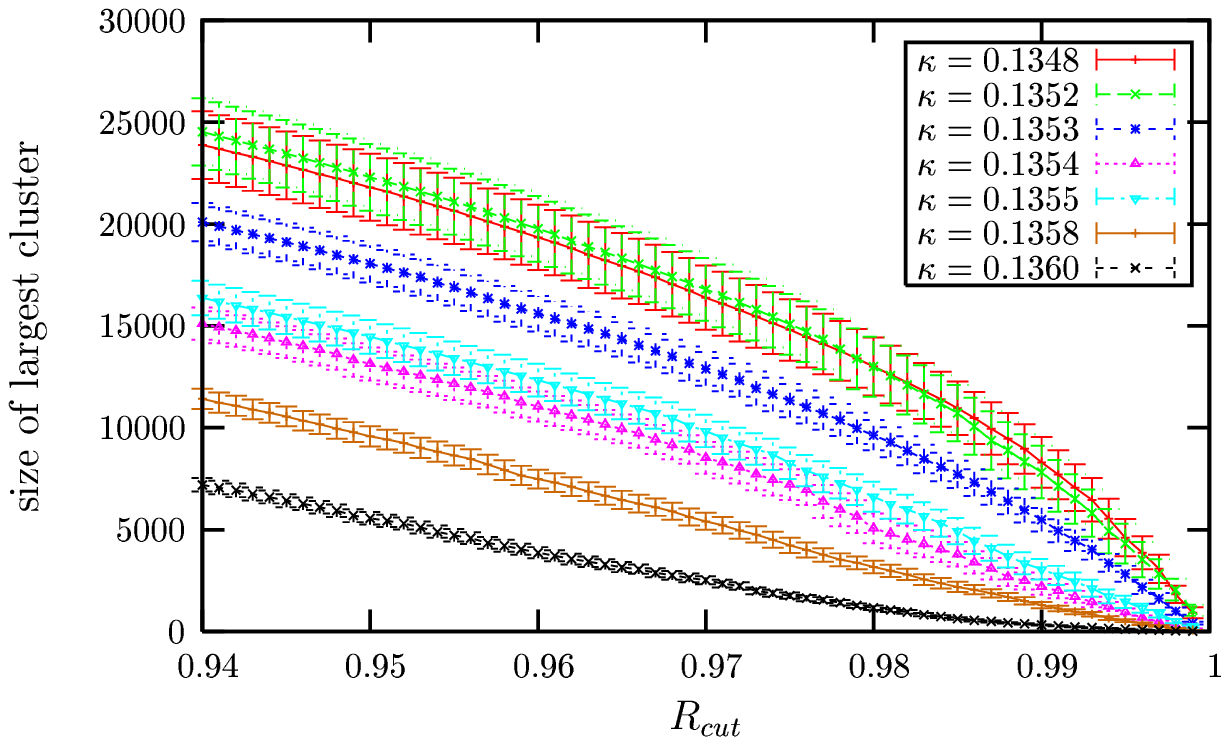}&
\hspace*{-0.5cm}\includegraphics[height=3.8cm]{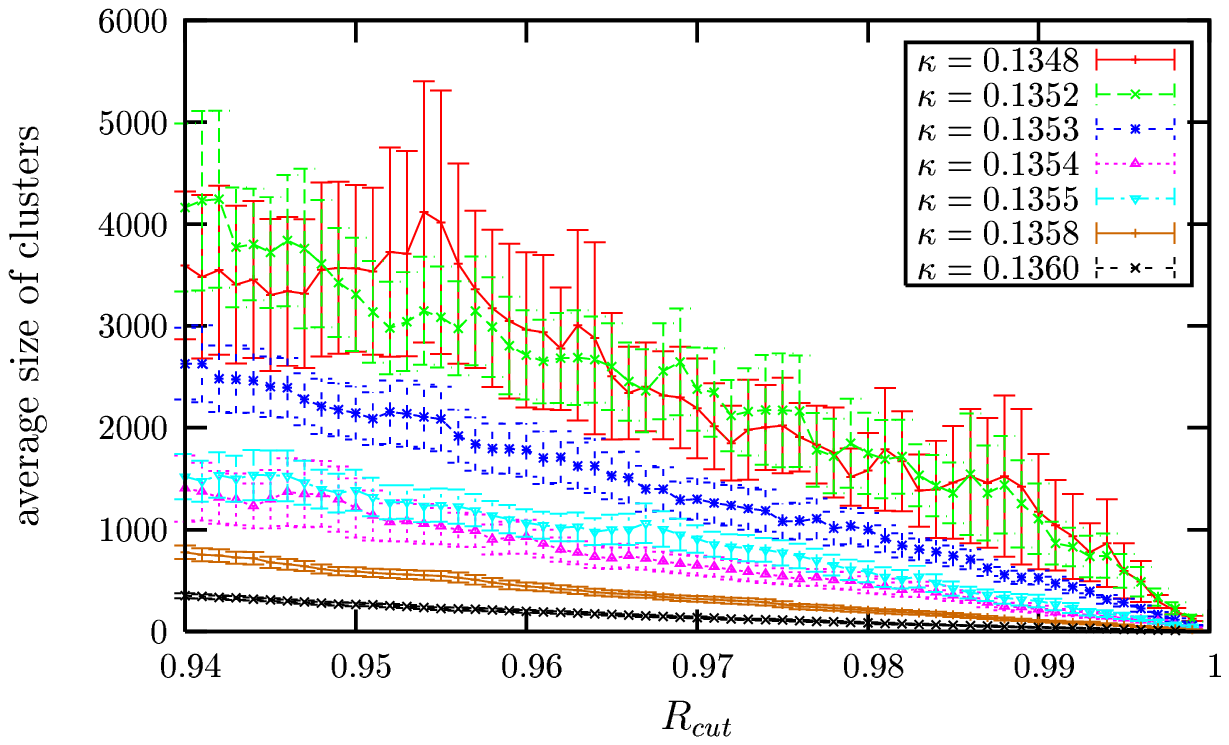}\\
(c) & (d) 
\end{tabular}
\caption{Results of the cluster analysis with respect to the local degree of
(anti-)selfduality $R(x)$: (a) the number of clusters, (b) their connectivity,
(c) the average size of the largest cluster and (d) the average size of all 
clusters, all as function of the cutoff $R_{cut}$.}
\label{fig:rcluster}
\end{figure}
\end{center}

\vspace*{-1cm}

Traditionally, the chiral symmetry restoration has been explained by pairing
of instantons and antiinstantons. To describe the changes in the (anti-)selfdual  
structure in more detail and in a model-independent manner,
we perform a cluster-analysis 
with respect to $R(x)$. Fig.~\ref{fig:rcluster} (a) shows the number of clusters 
consisting of link-connected sites $x$ with $|R(x)| \ge R_{cut}$ 
as a function of the lower cutoff $R_{cut}$. One can see that 
for $\kappa<0.1355$ in the confinement phase the number of clusters 
is surprisingly stable with increasing $R_{cut}$, whereas for the 
largest two $\kappa$ values in the deconfined phase the number of clusters 
decreases rapidly towards large cutoffs, {\it i.e.} a high degree of 
(anti-) selfduality. In Fig.~\ref{fig:rcluster} (b) we show the connectivity 
$f(r_{max})$ of these clusters, {\it i.e.} the probability for two 
lattice points, separated by the maximal possible distance, to belong to the 
same cluster. Generally one can see that the larger the $\kappa$ values are, 
the smaller this probability is. For the largest two $\kappa$ values, percolation 
({\it i.e.} $f(r_{max})>0)$ completely disappears at $R_{cut} \approx 0.97$.  Clusters that are more (anti-)selfdual than
that are well isolated. 
On the other hand, in the low-temperature phase 
percolation 
exists almost up to $R_{cut}=0.999$, indicating that perfectly 
(anti-)selfdual objects penetrate throughout the whole lattice volume.
The average size of the largest cluster and the average size of all clusters, 
shown in  Fig.~\ref{fig:rcluster} (c) and (d), respectively, as function of 
$R_{cut}$ strongly decreases with higher temperatures. 

\newpage

\vspace*{-1.5cm}
\section{Topological properties in the vicinity of the phase transition}

\begin{figure}[t!!]
\begin{center}
\hspace*{-0.5cm}\begin{tabular}{cc}
\epsfig{file=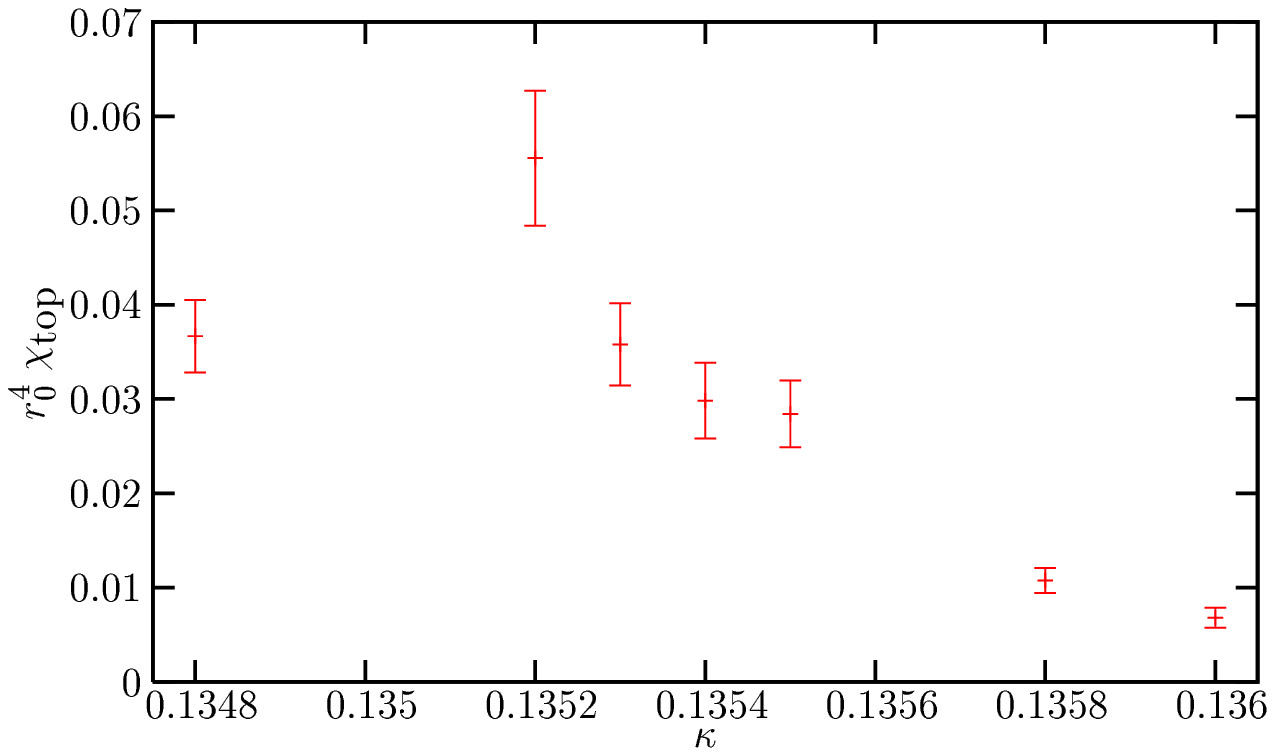,height=4cm}&
\epsfig{file=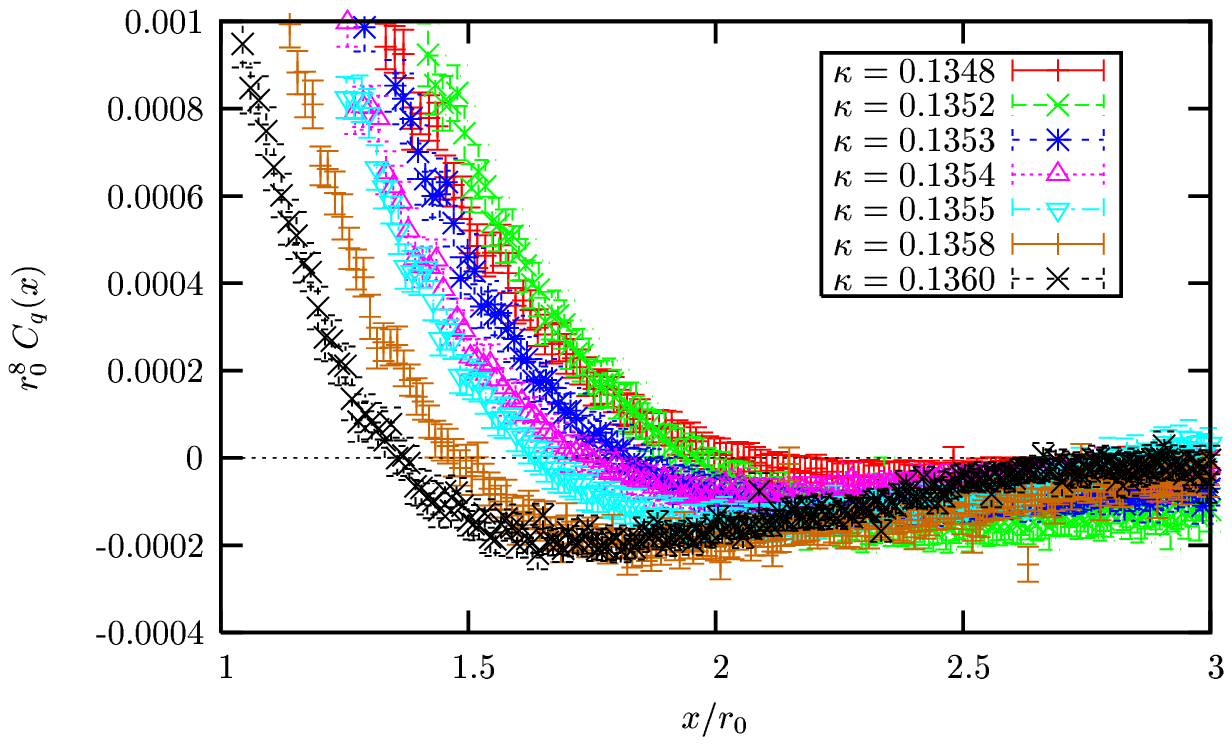,height=4cm}\\
(a) & (b)
\end{tabular}
\caption{ 
(a) The topological susceptibility $\chi_{top}$ vs. $\kappa$.
(b) The correlator $C_q(x)$ of the UV-filtered topological charge density 
computed from the lowest 50 eigenmodes. The plot focusses on the 
region where the correlator turns negative.}
\label{fig:topology}
\end{center}
\vspace*{-0.4cm}
\end{figure}

Since overlap fermions offer an exact realisation of the Atiyah-Singer index 
theorem at finite cutoff $a$, the global topological charge is given as 
$Q=\sum_{i \in {\rm~zero modes}} \sum_x p_{5~i}(x)$. 
The topological susceptibility $\chi_{top}=\langle Q^2\rangle/V$ 
obtained from this fermionic definition of $Q$ is displayed in 
Fig.~\ref{fig:topology} (a) and shows a rapid drop~\footnote{The data for 
the lowest two $\kappa$ values is likely to change with increased statistics. 
The $Q$-distributions (not shown here) for these 
two $\kappa$ values exhibit strong deviation from  Gaussian shape.} 
in the analysed temperature interval $[0.91~T_c,1.09~T_c]$.
The susceptibility can be expressed as the integral
$\chi_{top}=\int dx \; C_q(x)$ 
over the topological charge density correlator $C_{q}(x)=\langle q(0)q(x) \rangle$. 
Here we use the truncated eigenmode expansion of the topological charge density, 
$q_{IR}(x)=-\sum_{i=1}^n (1-\frac{\lambda_i}{2})\; p_{5~i}(x)$,    
including $n=50$ eigenmodes in the ``filter''.
The topological charge correlator $C_q(x)$  presented 
in Fig.~\ref{fig:topology} (b) shows a 
gradual change at the 
transition, revealing a short range charge compensation (traditionally interpreted as instanton-antiinstanton pairing) in the high-temperature phase.

\vspace*{-0.3cm}
\section{Summary}

We have complemented the efforts of the DIK collaboration to locate the 
confinement / deconfinement  transition 
using gluonic signals (the Polyakov loop susceptibility) 
by a fermionic approach using valence overlap fermions as a probe. 
We observe that the chiral susceptibility shows a peak at a value
$\kappa\approx 0.1352$ which is lower than $\kappa_t=0.13542(6)$ as determined by the Polyakov loop 
method. On the other hand, 
a gap in the spectrum does not open below $\kappa \approx 0.1358$. 
The opening of the gap is preceded by the low-lying modes becoming more and 
more localised.  
When the gap opens (apart from a few modes in the sample falling into the gap, which exclusively belong to the first pair of nonzero modes) 
the local chirality of the near-zero modes and the (anti-)selfduality which they contribute to the field strength tensor  is almost completely lost. Only below 
$\kappa \approx 0.1358$ extended, approximately (anti-)selfdual domains 
(say with $R_{cut} \gtrapprox 0.97$) percolate throughout the whole lattice 
volume. The disappearance of such extended structures in the high-temperature 
phase is also reflected in the topological charge correlator, which signals 
some short-range charge compensation in the high-temperature phase.
Thus it seems that different observables used to locate the transition  yield different critical values for  $\kappa_t$.
This could be  a hint that in $N_f=2$ clover-improved QCD the finite temperature transition is 
realised as a  crossover where various transition
phenomena take place~\cite{Aoki:2006we}.
The nature of the real transition to the quark-gluon plasma 
remains an open
question and requires further investigation.

\subsubsection*{Acknowledgements}
The numerical overlap calculations have been performed on the IBM p690 at HLRN
(Berlin). Part of this work is supported by DFG under contract FOR 465.


\begin{thebibliography}{99}

\bibitem{fodorandkarschtalk} Z.~Fodor, \pos{PoS(LATTICE 2007)011}; 
F.~Karsch, \pos{PoS(LATTICE 2007)015}.


\bibitem{Bornyakov:2004ii}
  V.~G.~Bornyakov {\it et al.}  [DIK Collaboration],
  %``Finite temperature QCD with two flavors of non-perturbatively improved
  %Wilson fermions,''
{\em Phys.\ Rev.}   {\bf D71}, 114504 (2005)
 [\href{http://arXiv.org/abs/hep-lat/0401014}{{\tt hep-lat/0401014}}].


\bibitem{Bornyakov:2005dt}
  V.~G.~Bornyakov {\it et al.}  [DIK Collaboration],
  %``Critical temperature in QCD with two flavors of dynamical quarks,''
  \pos{PoS(LAT2005)157}.

\bibitem{bornyakovtalk} V.~G.~Bornyakov {\em et~al.} [DIK Collaboration], \pos{PoS(LATTICE 2007)171}.

\bibitem{Weinberg:2005dh}
 V.~Weinberg  {\it et al.},
  %``Probing the chiral phase transition of N(f) = 2 clover fermions with
  %valence overlap fermions,''
 \pos{PoS(LAT2005)171}.

\bibitem{Ilgenfritz:2007xu}
 E.-M.~Ilgenfritz  {\it et al.},
  %``Exploring the structure of the quenched QCD vacuum with overlap fermions,''
{\em Phys.\ Rev.}   {\bf D76}, 034506 (2007),
\href{http://arXiv.org/abs/0705.0018}{{\tt 0705.0018 [hep-lat]}}.




\bibitem{ilgenfritztalk} E.-M.~Ilgenfritz {\em et~al.}, \pos{PoS(LATTICE 2007)311}.

\bibitem{Horvath:2001ir}
  I.~Horvath  {\em et~al.}, 
 %{\it Evidence against instanton dominance of topological charge fluctuations  in QCD},
  {\em Phys.\ Rev.}  {\bf D65}, 014502 (2002), 
  [\href{http://arXiv.org/abs/hep-lat/0102003}{{\tt hep-lat/0102003}}].

\bibitem{Gattringer}
  C.~Gattringer,
 % {\it Testing the self-duality of topological lumps in SU(3) lattice gauge theory},
{\em  Phys.\ Rev.\ Lett.} {\bf 88}, 221601 (2002),  
 [\href{http://arXiv.org/abs/hep-lat/0202002}{{\tt hep-lat/0202002}}].

\bibitem{Aoki:2006we}
 Y.~Aoki  {\it et al.},
  %``The order of the quantum chromodynamics transition predicted by the
  %standard model of particle physics,''
  Nature {\bf 443} (2006) 675,
 [\href{http://arXiv.org/abs/hep-lat/0611014}{{\tt hep-lat/0611014}}].


\end{thebibliography}
\end{document}